\documentclass[10pt,a4paper,twocolumn]{article}
\usepackage[utf8]{inputenc}
\usepackage{amsmath}
\usepackage{amsfonts}
\usepackage{amssymb}
\usepackage{graphicx}

\usepackage[backend=bibtex,style=numeric,sorting=none]{biblatex}

\usepackage{authblk}

\usepackage{hyperref}

\bibliography{chgcnn}

\makeatletter
\renewcommand{\@seccntformat}[1]{}
\makeatother

\usepackage{fullpage}

\title{Crystal Hypergraph Convolutional Networks}
\author{
Alexander J. Heilman, Weiyi Gong, and Qimin Yan
  }

\affil{\textit{
Department of Physics}\\
\textit{Northeastern University, 
360 Huntington Ave, Boston, MA 02115}}

\begin{document}
\twocolumn[
\maketitle
  \begin{@twocolumnfalse}
    \maketitle
    \begin{abstract}
    Graph representations of solid state materials that encode only interatomic distance lack geometrical resolution, resulting in degenerate representations that may map distinct structures to equivalent graphs. Here we propose a hypergraph representation scheme for materials that allows for the association of higher-order geometrical information with hyperedges. Hyperedges generalize edges to connected sets of more than two nodes, and may be used to represent triplets and local environments of atoms in materials. This generalization of edges requires a different approach in graph convolution, three of which are developed in this paper. Results presented here focus on the improved performance of models based on both pair-wise edges and local environment hyperedges. These results demonstrate that hypergraphs are an effective method for incorporating geometrical information in material representations.
    \end{abstract}
  \end{@twocolumnfalse}
  \vspace{.5cm}
]

\section{Introduction}
Machine learning has proven to be a computationally cost-effective and powerful predictive tool in the screening of large sets of material systems for certain material properties \cite{mlcite1, mlreview1, mlreview1.5, mlreview2, mlreview2.25, mlreview2.5, mlreview2.75, mlreview3}. Some of the most effective state-of-the-art models applied to invariant target predictions represent material systems as graphs \cite{schnet, cgcnn, megnet, chemgnn, geocgcnn, icgcnn}. These graphs encode physical properties in feature vectors associated with graph components, and update or 'learn' these features with a trained graph neural network or message passing network \cite{mpnn}. 

One problem with such graphical representations, however, is the lack of representation of higher-order geometrical structure, since the constructed crystal graphs can only include pair-wise descriptors. This may make it hard or impossible for models to distinguish between compositionally similar but structurally distinct systems with unique material properties \cite{congn}. Other works have approached this problem by including higher-order geometrical features such as overlapping bonds' angles \cite{alignn,m3gnet,congn}. However, these approaches come with a quadratic increase in the total number of messages with respect to $\Bar{N}_{\text{edges}}$, the average number of edges per atom.

Here, we propose the concept of \textit{crystal hypergraphs} to ail this lack of geometrical information in the more restrictive graph representations. In a crystal hypergraph, we may define larger (than strictly pair-wise) hyperedges that correspond to higher-order geometrical structures of material systems explicitly, such as triplets of neighboring atoms, or coordination polyhedra/motifs \cite{paulings_rules, coordpolyshapes, motifstats, motifexplore, clustermotifsanal, motife3nn}. These different structures then may have different coordinate invariant features associated with them, such as angles and local order parameters \cite{orderparam1, orderparam2, molorderparam}, respectively. Note that in that regard, crystal hypergraphs are naturally heterogeneous in their hyperedges, since there are different feature sets for different types of hyperedges.

Of course, the definition of a more general hypergraph representation requires the generalization of the message passing framework mentioned above. Here, we propose three possible approaches to such a generalization that handle the now-variable size of hyperedges. In a certain sense, these allow for the learning of a certain type of 'cluster-correlation expansion' \cite{clease, cce_crys, cce_gen} by the model, where clusters of interest correspond to the hyperedges defined.

As a proof of concept, we propose and implement a crystal hypergraph convolutional model (CHGCNN) that incorporates invariant geometric features for bonds, triplets, and motifs of crystals as hyperedge features. This allows us a unique opportunity to demonstrate the importance of different order structures for these different material property prediction tasks. Namely, we compare the performance of models based on atom, bond and triplet information against those incorporating atom, bond, and motif information (i.e. first shell hyperedges) on various predictive tasks with varying data sizes.

Results presented here indicate that first-shell (motif) hyperedges may be sufficient, if not more informative, than triplet hyperedges for many common predictive tasks. This comes at a substantially lower computational cost, in terms of the total number of messages exchanged through graph convolution.

The structure of this work then is as follows: first, we give a brief overview of crystal graph construction and message passing networks. A motivating representation problem is then identified with our definitions and the concept of crystal hypergraphs is introduced, with a particular focus on different types of hyperedges and their corresponding feature sets. Three generalized message passing frameworks are then considered, and a specific model architecture is presented. Finally, this specific architecture is used on various datasets to compare performance of different sets of hyperedge types.

\section{Crystal Graphs}
A common representation of crystalline systems in machine learning is via graphs (that is, collections of nodes and binary connections between them). We may define a crystal graph $\mathcal{G}=\lbrace \mathcal{V},\mathcal{E}\rbrace$ as a set of vertices $v_i\in\mathcal{V}$, corresponding to each atom $i$, and edges $e_{ij}\in\mathcal{E}$, where edges are determined by some physical criteria.

Physical information is then associated with the objects in these graphs by way of feature vectors. These are vectors with components describing the physical characteristics of their corresponding graph component, and which may be further 'learned' or updated through a graph neural network.


A commonly applied criteria for the formation of edges between atoms is a combination of a maximum distance cutoff $r_{max}$ and a maximum number of neighbors for each node $N_{max}$. That is, for each atom, edges are constructed between itself and it's $\leq N_{max}$-th closest neighbors in the crystalline structure within a shell of radius $r_{max}$.




The nodes' feature vectors encode the atomic information of the sites they describe. Two usual techniques include: explicitly engineered feature vectors (as in \cite{cgcnn}); and the learning of encodings for atomic sites based only on their atomic number (as in \cite{megnet}), beginning with some random initialization. Edge features are often derived exclusively from their distance.


Crystal graphs are usually constructed solely for use in some graph convolutional neural network. Perhaps the most general framework in which we may define graph convolution is the message passing framework, defined by Gilmore \cite{mpnn}.
A message passing network updates nodes based on 'messages' generated by the features of, and passed through, neighboring nodes (that is, nodes sharing an edge).



\begin{figure*}
    \centering
    \includegraphics[scale=0.63]{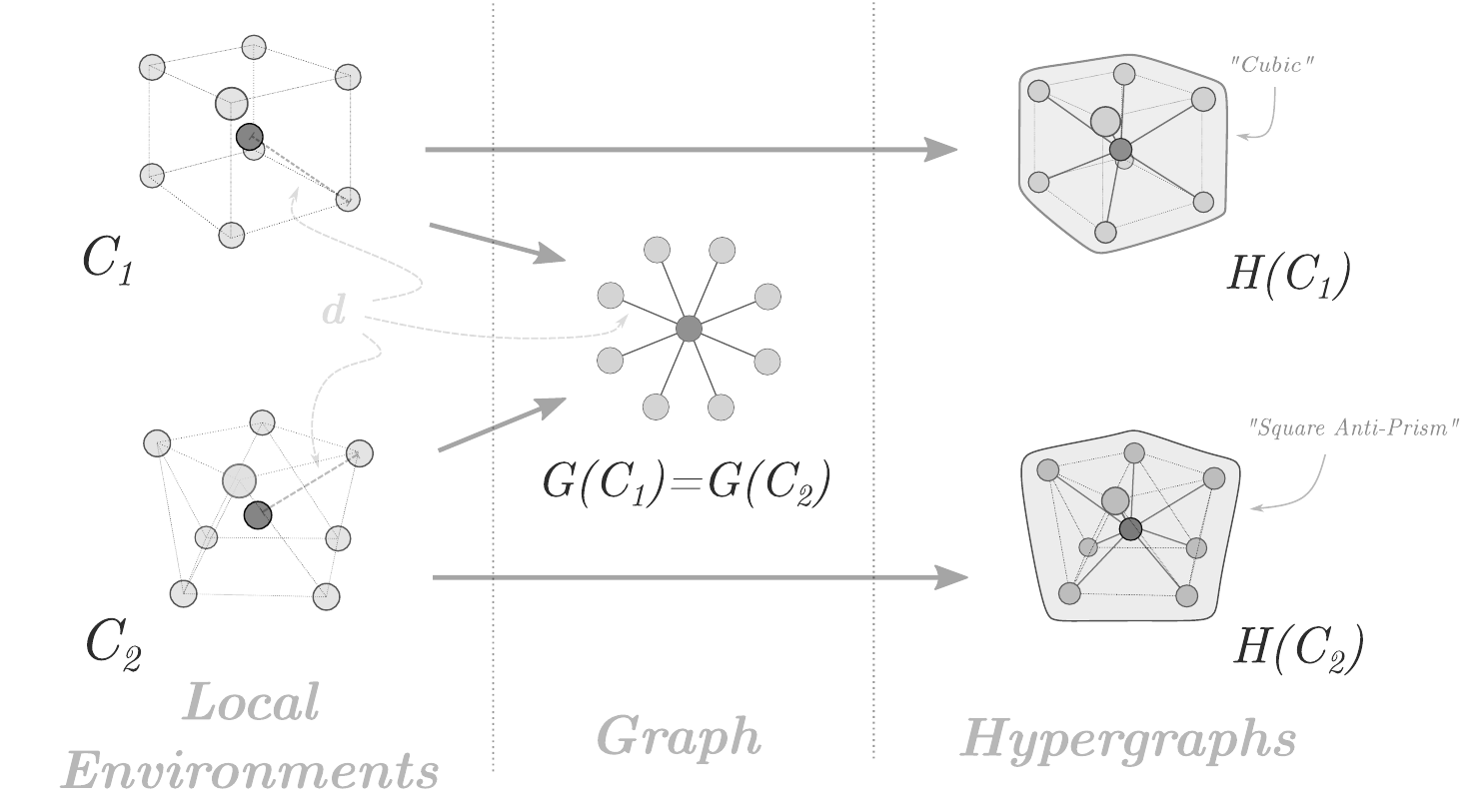}
    \caption{An example of two distinct geometries that are mapped to the same distance-based crystal graph. With inclusion of a first-shell feature vector encoding local geometry however, these structures are mapped to two distinct crystal hypergraphs. Note these are two possible coordination environments in oxides, determined statistically in \cite{motifstats}.}
    \label{fig:graph_cntex}
\end{figure*}

The construction above lacks higher-order geometrical information, i.e. local geometrical environments of atoms (that is, motifs) and global crystalline symmetries.
As a simple example of the low resolution manifest in crystal graphs, consider two atomic systems below: one with a local cubic symmetry, and another with a square anti-prism local environment; but both with the same bonding atoms. As demonstrated in Figure \ref{fig:graph_cntex} both structures would map to the exact same crystal graph, but could be easily be distinguished with an additional descriptor describing the local geometry of each's central atom.

Alternatively, one could include atomic position in the node features or a vector direction in edge features. However, this generally requires a unique treatment of such coordinate-system dependent information through convolution if the output is to maintain \textit{invariance} with respect to changes in coordinate system. As such, often only coordinate system invariant features are included in crystal graph representations, such as distance and atomic properties. 

\section{Crystal Hypergraph Construction}
The method proposed here solves the above problem by allowing for the explicit incorporation of this higher-order geometrical information in the form of hyperedges, which can be used to directly represent these higher-order structures.

A crystal hypergraph $\mathcal{H}=\lbrace\mathcal{V}, \mathit{H} \rbrace$ is a collection of nodes $v_i\in \mathcal{V}$ and hyperedges $h_j\in \mathit{H}$ (containing an arbitrary number of nodes), where the hyperedges are most generally heterogeneous. That is, we may wish to describe different types of hyperedges (e.g. bonds, triplets, and motifs) in the same hypergraph. These objects then have associated feature vectors encoding relevant physical information, which we also refer to as $v$ and $h$.

For the purpose of modeling material systems, we need to identify what different order structures are most important in their representation. Of course, atomic and bond level information is particularly important. However, higher-order structures may also be of interest, such as: triplets of atoms and local environments of atoms, which we refer to as motifs in crystals.

Each of the aforementioned structures also has a natural set of distinct, coordinate-system invariant features that may be associated with them. At the triplet level (where two bonds share some common node), there is always a corresponding angle. While at the motif level, order parameters \cite{orderparam1, orderparam2} or continuous symmetry measures \cite{csm_polyhedra, chemenv} may be used to describe 3 dimensional coordination environments quantitatively. 
These different order structures may all be represented in a single crystal hypergraph.

Below, we discuss 
the generation of, and association of features with, all of the above mentioned structures in crystalline solids. 

\begin{figure*}[!ht]
    \centering
    \includegraphics[scale=0.26]{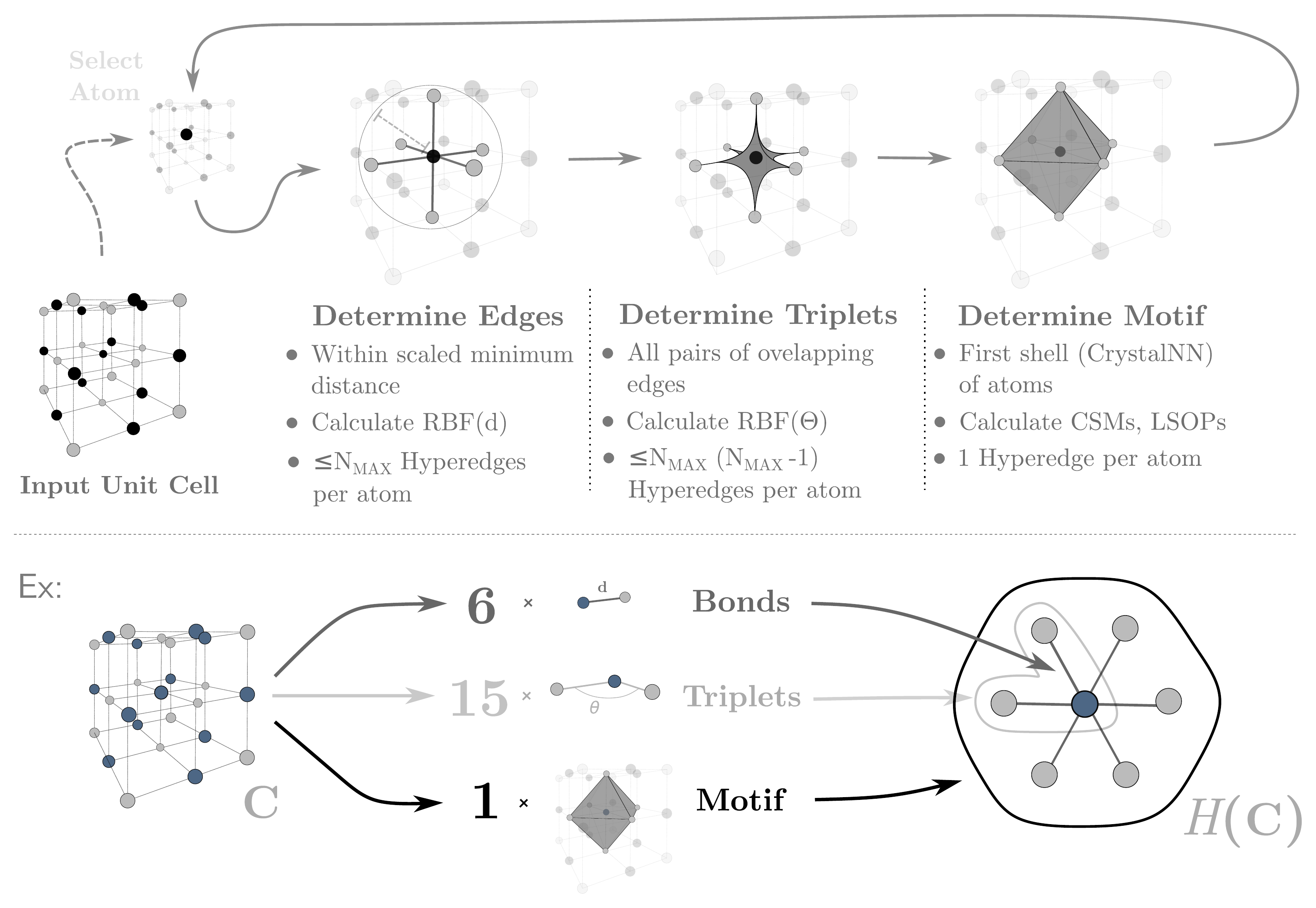}
    \caption{Typical construction loop for a crystal hypergraph. First, pair-wise bonds/edges are determined, then triplets are derived from overlapping pairs of bonds, and finally motifs are determined as first-shells of neighbors by some (generally more restrictive) criteria. Features for each and upper bounds on numbers of hyperedges for each type are also listed.}
    \label{fig:hypergraph-loop}
\end{figure*}

\subsection{Bond Edges}
Bonds, or pair-wise atomic connections, are determined in the same manner as in a crystal graph. In the results below, we choose edges from a maximum number of neighbors $N_{max}=12$ found within a shell of radius $r_{max}=6\AA$. 

\subsection{Triplet Hyperedges}
Triplet hyperedges are then formed from the set of bonds. For each set of bonds connected by one node, a triplet hyperedge is formed. 

The feature of these triplet hyperedges is also a Gaussian expansion, though now of the angle formed by the unit vectors of the two bonds \cite{alignn}.
Triplet hyperedges give us a way to incorporate some angular resolution into our representation scheme in a  coordinate-system invariant manner.
For a node with $N$ bonds then, there will be $N(N-1)/2$ triplets. Thus, the price we pay for complete angular resolution of any two bonds is a quadratic increase in the number of hyperedges.





\subsection{Motif Hyperedges}
Motif determination may be achieved by a wide range of functions, and is akin to an algorithmic determination of coordination number \cite{coordination_comp}. Here, we use a modified Voronoi algorithm with a cut-off radius implemented as CrystalNN in pymatgen.
Note this is a much stricter algorithm than that used to determine edges and triplets, since the motifs features depend heavily on the selected first-shell.

The features of these motifs are a concatenation of Zimmerman's 35 local structure order parameters \cite{orderparam1,orderparam2}, and continuous symmetry measures \cite{csm_polyhedra} (e.g. 'distance to a perfect shape') for 59 common coordination environments. In essence, both are just sets of quantitative measures designed to describe 3 dimensional physical shape. 
Motifs give us a way to describe the local geometry of sites in material systems with much fewer hyperedges. Since each node will contribute one motif hyperedge, for a crystal with $n$ nodes, we just have $n$ motifs.

\section{Crystal Hypergraph Convolution}

We now must consider a message passing framework analogous to \textit{Gilmore, et al} \cite{mpnn} but applying to hypergraph structures. That is, we now have:

\begin{align*}
m_v^{t+1}&=\sum_{h_j\in \mathcal{N}(v)} M_t(n_v^{t},h_j^{t},\lbrace n_w^t \vert n_w \in h_j \rbrace),\\
n_v^{t+1}&=U_t(n_v^t,m_v^{t+1}),\\
\hat{y}&=R(\lbrace n_v^T\rbrace),
\end{align*}
so that each node is still updated according to layer-wise update function $U_t$, aggregating messages $m^{t+1}$ formed from origin node features, hyperedge features $h_j$, and hyperedge neighborhood features $n_w \in h_j$, in analogy to the graph-based MPNN approach. This updating occurs node-wise and then after $T$ layers, some readout function $R$ is again used to output the corresponding predicted value $\hat{y}$, which utilizes the set of learned node features.

  
The biggest difference here is that we now need a message forming function $M_t$ that accounts for a set of node features $\lbrace n_w^t \vert n_w \in h_j \rbrace$, that may vary in size between different hyperedges (even of the same type). This stands in opposition to the case of regular edges, where we are assured a fixed size of two nodes per edge. 

One approach would be to fix the dimensionality of each type of hyperedge, or have a different convolutional operator for each different size hyperedge (as is effectively the approach taken with line graph networks \cite{linegraph_general}). Here, however, we wish to maintain generality in edge size so we need not fix hyperedge sizes for hyperedge type, since structures of different sizes may be described by similar metrics. For example, motifs resembling polyhedra with different numbers of vertices may be described by common sets of features.



Of course, there should be different message and update functions for each different order structure (bonds, triplets, motifs, etc.) with different features. This is accounted for by treating the data as a heterogeneous graph, with different hyperedge types. Below, we consider three strategies that allow us to apply our convolutional operator to hyperedges of arbitrary size.

\subsection{Three Possible Approaches to Hyperedge Convolution}

Three general approaches for message passing that account for this multi-order nature have been considered in this work: \textbf{1.} the construction of a hyperedge relatives graph, upon which regular graph convolution may be applied; \textbf{2.} total exchange hyperedge message convolution, which completely generalizes the CGCNN \cite{cgcnn} and ALIGNN \cite{alignn} models to hypergraphs; and \textbf{3.} neighborhood aggregation, which balances performance of the former approach by forming a single neighborhood feature for each hyperedge.

Each approach has a different computational cost in terms of total number of messages, along with a potentially different practical definition of a hypergraph. These considerations are presented below, with a specific convolutional structure and empirical results on common test datasets then presented after.


\subsubsection{Hyperedge Relatives Graph}
We may define a dual graph $\mathcal{D}(h)$ to a hypergraph $h$ to be a graph in which nodes represent the hyperedges of the hypergraph, and connections represent the overlap of respective hyperedge neighborhoods. 
In the case of a crystal hypergraph with heterogeneous hyperedges, this dual graph is a graph with heterogeneous nodes. We term this heterogeneous dual graph of a crystal hypergraph the relatives graph for simplicity. 
Atomic features may be included in this framework by adding a singleton hyperedge for each node.

The definition of the relatives graph allows us to perform the usual methods of graph convolution on hyperedge features. Such an approach also allows us to define our relatives graph as we would a graph, with just a standard edge index.

\begin{figure*}
    \centering
    \includegraphics[scale=.7]{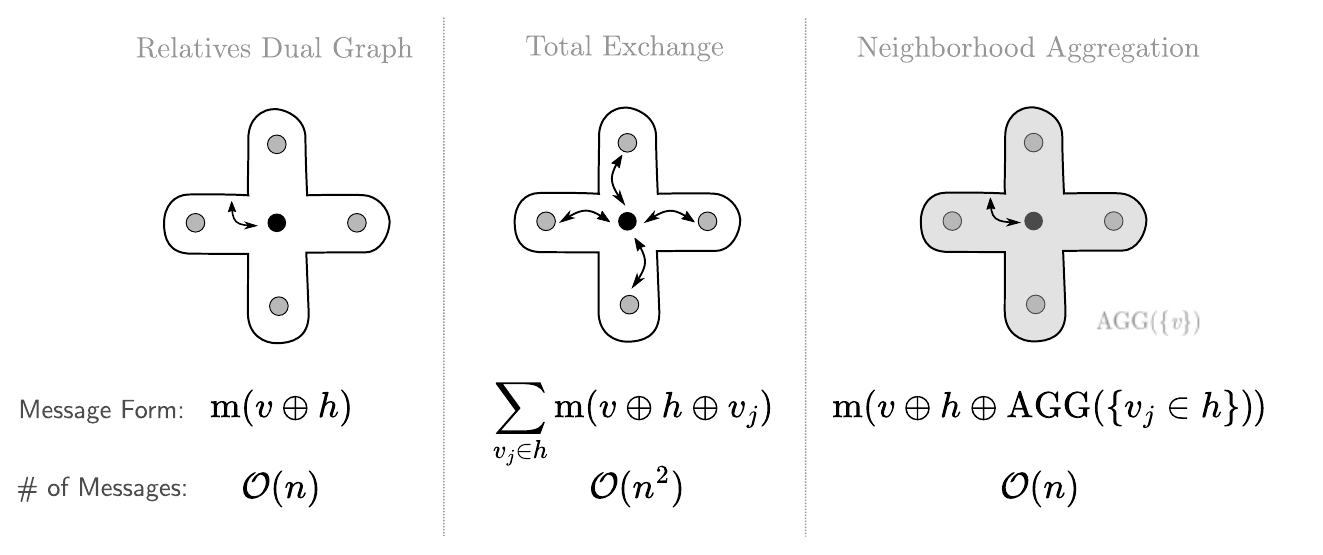}
    \caption{Overview of three possible message functions $m$ for nodes $v$ that generalize the message function in \cite{mpnn} to hyperedges $h$ with more (or less) than two nodes. Here, $n$ represents the average number of nodes in a hyperedge and the scaling relation is then for the total number of messages for one layer, for each hyperedge, for each approach.}
    \label{fig:hmpnn}
\end{figure*}

However, this approach lacks the interaction of neighboring features in convolution via the connecting hyperedge. That is, without a clear definition of the edge attribute, messages are generally of the form below:
$$
m_v^{t} = \sum_{h_j\in \mathcal{N}(v)}M_t(n_v^{t},h_j^{t})
$$
in which we simply discard the neighborhood of other node features contained in the hyperedge.

Computationally, this approach has a total number of messages that scales linearly with average hyperedge size, since each hyperedge only contributes one message to each node it contains.
Accounting only for node-hyperedge connections in a relatives graph derived from a hypergraph with $m$ hyperedges of average order $n$, the total number of messages per convolution will scale as $\mathcal{O}(nm)$.
 
\subsubsection{Total Exchange Message Passing}
Of course, we may wish to incorporate the neighboring features of some representation via their connecting hyperedge. This may be accomplished by simply forming a message for every pair of connected representations along with their connecting hyperedge's representation.
$$
m_v^{t} = \sum_{h_j\in \mathcal{N}(v)} \sum_{n_w \in h_j } M_t(n_v^{t},h_j^{t}, n_w^t),
$$
Here, though, we've introduced a new summation which may drastically increase the number of messages for larger hyperedges. In this scheme, if each hyperedge contains an average of $n$ nodes and there are $m$ hyperedges, the total number of messages exchanged per node-wise convolution will scale as $\mathcal{O}(n^2m)$.


\subsubsection{Neighborhood Aggregation}
Since the number of messages will scale tremendously with larger hyperedges in the framework described above, we may seek a way to incorporate the neighborhood of features of a hyperedge into a single message.

In this case, we may essentially form a 'neighborhood feature' representative of all a hyperedge's contained nodes. Typical aggregation methods may be used and trained to perform this neighborhood feature generation. Here then, we deal with message functions of the form:
$$
m_v^{t+1}=\sum_{h_j\in \mathcal{N}(v)} M_t(n_v^{t},h_j^{t},\text{AGG}\big(\lbrace n_w^t \vert n_w \in h_j \rbrace\big)
$$
This results in a number of node-wise messages that scales linearly with the average size of hyperedges, so that we now have a relationship of order $\mathcal{O}(nm)$ again, while still incorporating the features of neighboring nodes.

\subsection{Example Model Architecture}
Initial atomic features were those used in CGCNN \cite{cgcnn}, consisting of a concatenated set of one-hot encoded atomic properties. For bond features, we used a Gaussian expansion of interatomic distance of dimension 40 ranging from 0 to 6 \AA, triplet features were a Gaussian expansion of the cosine of bond angle, also of dimension 40, and motif features were a concatenation of 94 scalar order parameters and continuous symmetry measures. In the model considered in this work, initial node and hyperedge features were first passed into a linear embedding layer (with no activation function) with an output dimension of 64. 

\begin{figure*}
    \centering
    \includegraphics[scale=.75]{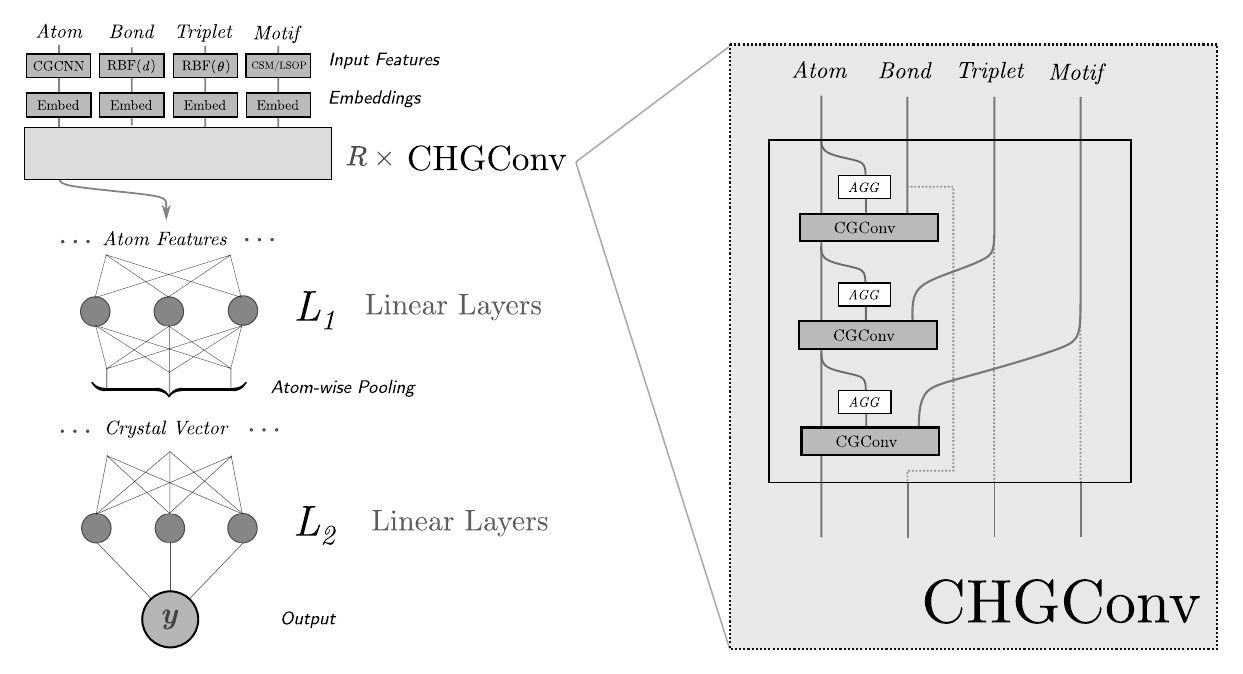}
    \caption{Example architecture for the crystal hypergraph convolutional network implemented in this work. Essentially, the model is a generalizaiton of CGCNN's \cite{cgcnn} model architecture with CGConv being replaced by $R$ hypergraph convolutional layers (CHGConv). Here, CHGConv updates nodes first according to edges (bonds), and then according to triplets or motifs. }
    \label{fig:architecture}
\end{figure*}

These embedded features were then fed into a set of Crystal HyperGraph Convolutional (CHGConv) layer which utilizes the neighborhood aggregation method presented above. In CHGConv, we use a set of CGConv \cite{cgcnn} layers applied to consecutively larger hyperedge types, taking as input the origin node of the hyperedge, the connecting hyperedge feature as the edge feature, and an aggregated set of neighborhood features as the connected node feature. So, for every CHGConv layer, the atoms are updated by each hyperedge type chosen once (see Fig. \ref{fig:architecture}). Note that each CGConv for different hyperedge types have independent trainable parameters. 

These learned node features are then mean pooled to form a crystal vector, which is passed to a fully connected layer and then projected down to a one-dimensional (scalar) output for regression. In the case of classification tasks, the fully connected layer, after mean pooling, utilized a dropout mechanism and output a probability distribution of classes via a softmax activation function.

\section{Training and Results}
Crystal hypergraph networks provide a unique opportunity to investigate the importance of different order structures in the prediction of various material properties. Specifically, we may compare performance between models based on different types of hyperedges to probe the relevance of certain structures (e.g. motifs vs. triplets) in material property prediction. From the different hyperedge types considered here, we build five different models based on the architecture given in Fig. \ref{fig:architecture}. We consider three models incorporating only one type of hyperedge: bond-only, triplet-only, and motif-only; as well as two models incorporating two types of hyperedges: bond-and-triplet, and bond-and-motif models. For compound models (including more than one hyperedge type) each CHGConv layer performs convolution over the hyperedges in ascending order of hyperedge size. These models were each trained for 300 epochs on sets of training data from two different databases of material properties: the Materials Project \cite{matproj} and MatBench \cite{matbench}. Training was then performed on 80\% of the data with 20\% withheld for validation.

We first focus on the comparative performance of models incorporating either motif-only, bond-only, or both bond-and-motif level hyperedges on three targets for 152,605 materials from the Materials Project database. Results for this set of tests are indicated in Table \ref{fig:mp_table}, with targets including formation energy, band gap, and metalicity.

For the Materials Project dataset, the bond-and-motif model performed best for all tasks. In the prediction of formation energy, the motif-only model performed better than the bond-only model with an MAE of 0.088 eV/atom vs. an MAE of 0.177 eV/atom, while the combined motif-and-bond model performed better than both with an MAE of 0.074 eV/atom on the validation set. For the prediction of band gaps, the bond-only model performed better with an MAE of 0.315 eV than the motif-only model with an MAE of 0.387 eV. However, the bond-and-motif model also performed best on the band gap dataset with an MAE of 0.301 eV on the validation set. This trend also held for the classification task of metal/non-metal, with the best performance on the validation set again by the bond-and-motif model with an accuracy of 86\%, vs. accuracies of 84\% and 85\% from the bond-only and motif-only models, respectively. This clearly suggests that motif-level hyperedges contribute significantly to the performance of hypergraph models.

We now compare performance on five MatBench datasets to compare performance between triplet-level hyperedges and motif-level hyperedges, with results for this set of tasks given in Table \ref{fig:matbench_table}. These five datasets consisted of the following targets: the highest frequency phonon peak for 1,265 materials, refractive indices for 4,764 materials, formation energies for a set of 18,829 perovskite materials , and 10,987 bulk and shear moduli.

On most tasks of this set, the largest models incorporating both bonds and higher-order hyperedges did perform the best. Except in the prediction of refractive indices, where motif-level information alone performed the best with an MAE of 0.485 on the validation set; though performance here was comparable to bond-only performance with an MAE of 0.497.
Triplet-and-bond models performed best overall on elastic tasks in the prediction of shear and bulk moduli $G_{vrh}$ and $K_{vrh}$, with MAEs on the validation of 0.088 Log$_{10}$(GPa) and 0.071 Log$_{10}$(GPa), respectively. This may be indicative of the importance of such information in the relation of stress to infinitesimal strain, since the initial angle formed by any two bonds would be of particular importance in considerations of shear response (that is, calculations of $G_{vrh}$) though, perhaps, less so in considerations of bulk response ($K_{vrh}$). This is corroborated by the difference in performance between motif and triplet models on both tasks, since motif information seems comparable in prediction of bulk moduli with an MAE of 0.073 Log$_{10}$(GPa) for the motif model (compared to 0.071 Log$_{10}$GPa for the triplet-based model), but less so in the prediction of shear moduli with an MAE of 0.095 Log$_{10}$(GPa) (compared to 0.088 Log$_{10}$GPa) . 
Motif-and-bond models performed best on the remaining tasks of highest frequency phonon mode peak, perovskite formation energy prediction, with MAEs of 64.5 cm$^{-1}$ and 0.0488 eV/atom, respectively.

\begin{table*}
\begin{tabular}{c|ccc}
Hyperedge & Form. Energy & Band Gap & Metal/Non-metal Classification \\
Types & Best MAE (eV/Atom) & Best MAE (eV) & Best Accuracy \\
\hline
Bond-only & 0.177 & 0.315 & 84\% \\
Motif-only &  0.088 & 0.387 & 85\%\\
Bond \& Motif &  \textbf{0.074} & \textbf{0.301 } & \textbf{86\%}\\
\end{tabular}
\caption{Validation dataset results for several Materials Project target sets. Here, each dataset included a total of 152,605 materials. Best results are indicated in bold.}\label{fig:mp_table}
\end{table*}

\begin{table*}\small
\begin{tabular}{c |ccccc}
 & Phonons  & Refractive Indices& Perovskites  & Log$_{10}$($G_{vrh}$) & Log$_{10}$($K_{vrh}$) \\
Hyperedge & (\textit{1,265}) &  (\textit{4,764}) & (\textit{18,829}) & (\textit{10,987}) & (\textit{10,987}) \\
Types & MAE (cm$^{-1}$) & MAE & MAE (eV/Atom)& MAE (Log$_{10}$GPa)& MAE (Log$_{10}$GPa)  \\
\hline
Bond-only & 84.1& 0.497 & 0.0584 & 0.099 & 0.083\\
Triplet-only & 71.3& 0.520 & 0.0566 & 0.094 & 0.073 \\
Motif-only & 77.6& \textbf{0.485} & 0.0611 & 0.103 & 0.077\\
Bond \& Triplet & 69.2& 0.550 & 0.0550 & \textbf{0.088}& \textbf{0.071} \\
Bond \& Motif & \textbf{64.5}& 0.510 & \textbf{0.0488}&0.095 & 0.073\\
\end{tabular}
\caption{Validation dataset results for several MatBench target sets, note that the italicized numbers below the target name correspond to the total size of each dataset. Best results are indicated in bold.}\label{fig:matbench_table}
\end{table*}

Perhaps the strongest point to be made in regards to these results is that for most tasks, motif information contributed to comparable or better performance than triplet-level results. This is at a significantly lower computational cost, in terms of the total number of messages exchanged through convolution, since the number of motifs is simply the number of atoms $n$, whereas the number of triplets scales with the average number of bonds per atom $N$ as $N(N-1)/2$.

A similar observation was made in the AMDNet architecture \cite{amdnet}, where motif information (included via an additional 'motif graph' for each material) also improved performance on most tasks, but here we compare results directly to the inclusion of bond angles via triplets.
Our results thus indicate that one local neighborhood feature per atom may be sufficient to describe the local geometries of atoms for many predictive tasks, as opposed to the more data-intensive triplet representation scheme usually employed (often by way of line graphs).  Taking this as a learned guiding principle, future crystal representations may benefit from reduced size while being assured similar geometric resolution.
Note that models using both motif and triplet-level edges, in general, diverged through training or did not perform any better than models using just motifs or triplets. As such, we only compare models using one or the other here.

State-of-the-art models applied to material property prediction often represent material systems as graphs with relatively low geometrical resolution. This low resolution is often increased by associating bond angles with auxiliary line graphs derived from the graph itself. The primary argument of this paper is that hypergraphs are a more natural representation of material structures that allow us to explicitly incorporate geometrical information with different substructures of our choice in one unified representation. Our results suggest that such an approach allows for a substantial decrease in computational cost by incorporating such geometrical information with single local environment hyperedges for each node as opposed to triplets of atoms for each pair of overlapping bonds. This is shown within one unified framework to have comparable performance on a number of common predictive tasks.

Future works may investigate more powerful hypergraph convolutional operators that automatically detect motifs \cite{contrastivelearn_motif, motifexplore, chemicalmotifcharacterization}; or apply this framework to molecular systems \cite{ molecule2, molecule3} with functional groups. Inter-order convolution may also be of interest for certain tasks, where different hyperedge types may update each other's representations as opposed to just atom representations. Note that inter-order convolution would allow for a complete generalization of previous line-graph convolution schemes, where triplets effectively update their respective bonds' representations through convolution, as in \cite{alignn, congn}. Other order structures (beyond motif-level) may also be of interest, such as hyperedges representing defect complexes or entire unit cells. Equivariant features and convolution \cite{e3nn, tfn, o3transformer1} may also be incorporated for the prediction of coordinate-system dependant properties of materials from hypergraph representations, with the present work being focused on coordinate-system invariant features and targets. 

\section{Acknowledgments}
This work is supported by the U.S. Department of Energy, Office of Science, Basic Energy Sciences, under Award No. DE-SC0023664. This research used resources of the National Energy Research Scientific Computing Center (NERSC), a U.S. Department of Energy Office of Science User Facility located at Lawrence Berkeley National Laboratory, operated under Contract No. DE-AC02-05CH11231 using NERSC award BES-ERCAP0029544.

\printbibliography

\medskip

\newpage

\appendix

\section{Motif Features: Structure Order Parameters \& Continuous Symmetry Measures}
The geometry of the motifs were incorporated as features composed of a concatenated list of structure order parameters and continuous symmetry measures (CSMs) for a set of common local environments. 

Structure order parameters are coordinate system invariant measures of 3 dimensional structure that are designed to be close to one when a given structure is similar to some prototypical arrangement. Note that this isn't in general a true 'distance'-like measure to some shape as a CSM is, however. The list of order parameters included those implemented in pymatgen code and described in \cite{orderparam1, orderparam2}.

A CSM is defined precisely so that it may act as a 'distance' from some prototypical shape to some given structure.

\section{CHGConv}\label{chgconv}
A specific implementation of a hypergraph convolutional operator in the hypergraph message passing framework is a generalization of CGConv implemented in pytorch geometric and based on CGCNN's convolutional operator defined in eq (5) of the original paper.

\begin{align*}
x_i^{t+1} &= \sum_{b_j} f(x_i^t, b_j,\text{AGG}(\lbrace x_j^t\in b_j \rbrace )) \\ 
& = \text{BN}\bigg[\sum_{b_j}\sigma \big(W_c\cdot [x_j\oplus b_j\oplus \text{AGG}(\lbrace x_j^t\in b_j \rbrace ] )\big)\\
&\quad\quad\cdot S^+ (W_f\cdot (x_j\oplus b_j\oplus \text{AGG}(\lbrace x_j^t\in b_j \rbrace ) )  ) \bigg]
\end{align*}
In the model utilized in this work, we generally employed use of a learnable set of common aggregation functions for the neighborhood feature aggregation ($\text{AGG}$ above), inspired by \textit{ChemGNN} \cite{chemgnn}.

\section{Hyperparameters for Testing}
For each convolutional structure, testing was done for a model with 3 convolutional layers. Each convolutional layer consists of back-to-back convolution from the smallest to the largest hyperedge type (for example two bond \& motif layers consist of a total sequence of bond, motif bond, motif). 

Stochastic gradient descent (SGD) was used as an optimizer through training with an initial learning rate of 0.01. A multi-step learning rate scheduler divided this learning rate by a factor of 10 at epochs 150 and 250, with training running for a total of 300 epochs. 

Hidden node features were of dimension 64 through all convolutional layers, and a hidden output layer of dimension 128 was used (similar to CGCNN's architecture). The loss functions utilized were MSE (for regression tasks) and cross entropy (for classification tasks). Accuracy is then reported in MAE for regression tasks and percentage correctly classified for classification tasks. Datasets were split 80\% for training and 20\% for validation tests.

\section{Comparison to Line Graph}
A more usual approach for the incorporation of bond angle information is via the construction of a line graph, as in \cite{alignn, m3gnet}. 
\begin{center}
\includegraphics[scale=0.38]{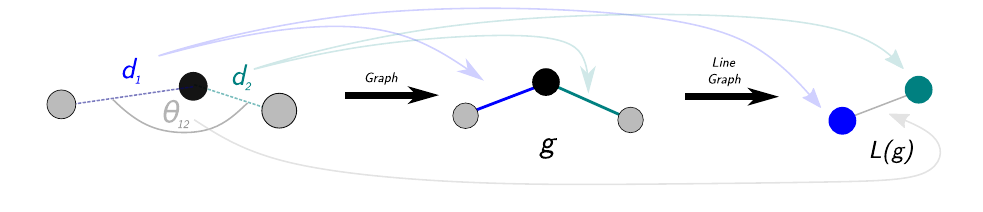}
\end{center}
These models generally first update the edge features of the crystal graph $\mathcal{G}$ by first applying some graph convolutional operator to the line graph $L(\mathcal{G})$ with angles encoded in  $L(\mathcal{G})$'s initial edge features.

Our argument against such representation schemes here is that the order of messages grows combinatorically for derived line graphs as $\mathcal{O}(nm^2)$, where $n$ is the number of nodes and $m$ is the average number of edges per node in $\mathcal{G}$.

Here, we incorporate a similar level of higher-order geometrical structure instead in a local environment, or 'motif', hyperedge (defined below). Note that these include only an extra number of messages on the order $\mathcal{O}(mn)$ if each node in a motif gets a message, or on the order $\mathcal{O}(n)$ if only center nodes are updated by their own motif hyperedges.

\section{Hyperedge Index}
Hypergraphs are treated as a set of node feactures $x$, hyperedge features $h$, and hyperedge indices $I$. 

The hyperedge index is, computationally, treated as a $[2,nm]$ dimensional vector (where $m$ is the number of hyperedges and $n$ is the avereage number of nodes contained in any hyperedge). 
The first index is the node contained and the second index is the containing hyperedge (as in \cite{hypergraphconv}).
\begin{center}
\includegraphics[scale = 1]{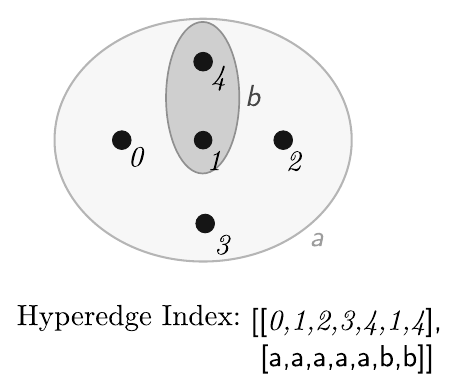}
\end{center}

\section{Code Availability}
The code used in this paper's results were built on pytorch-geometric and can be found in the following github repository: \hyperlink{https://github.com/qmatyanlab/CHGCNN}{https://github.com/qmatyanlab/CHGCNN}.

\end{document}